# Dense network of one-dimensional mid-gap metallic modes in monolayer MoSe$_2$ and their spatial undulations

Hongjun Liu[1], Lu Jiao[1], Fang Yang[2], Yuan Cai[3], Xianxin Wu[4,1], Wingkin Ho[1], Chunlei Gao[2], Jinfeng Jia[2], Ning Wang[3], Heng Fan[4], Wang Yao[1], and Maohai Xie[1*]

[1]*Physics Department, The University of Hong Kong, Pokfulam Road, Hong Kong*

[2]*Key Laboratory of Artificial Structures and Quantum Control (Ministry of Education), Department of Physics and Astronomy, Shanghai Jiaotong University, 800 Dongchuan Road, Shanghai 200240, China*

[3]*Physics Department, Hong Kong University of Science and Technology, Clear Water Bay, Kowloon, Hong Kong*

[4]*Institute of Physics, Chinese Academy of Sciences, Beijing 100090, China*

**We report the observation of a dense triangular network of one-dimensional (1D) metallic modes in a continuous and uniform monolayer of MoSe$_2$ grown by molecular-beam epitaxy. High-resolution transmission electron microscopy and scanning tunneling microscopy and spectroscopy (STM/STS) studies show these 1D modes are mid-gap states at inversion domain boundaries. STM/STS measurements further reveal intensity undulations of the metallic modes, presumably arising from the superlattice potentials due to moiré pattern and the quantum confinement effect. A dense network of the metallic modes with high density of states is of great potential for heterocatalysis applications. The interconnection of such mid-gap 1D conducting channels may also imply new transport behaviors distinct from the 2D bulk.**

*Email: mhxie@hku.hk

Layered transition metal dichalcogenides (TMDs), with the common formula of $MX_2$ (M=Mo, W; X=S, Se), are of great current interests for their potential electronic, optoelectronic and catalytic applications [1-3]. Remarkable intrinsic properties of these two-dimensional (2D) crystals have been discovered recently by optical and transport experiments [1,4-14]. At smooth edges or domain boundaries of the TMD monolayers (MLs), one-dimensional (1D) metallic states are found in the bulk gap, which have also attracted considerable interests [15-32]. These mid-gap states can affect the optical, transport and magnetic properties of the 2D materials [30-32]. Owing to their large density of states (DOS), the mid-gap metallic modes in TMDs are catalytically active for hydrodesulfurization (HDS) and hydrogen evolution reaction (HER) and are actively explored for catalytic applications [16-29]. To improve catalytic efficiency, networks of the edge sites have been widely adopted for HDS and HER applications [21,26,29].

In this Letter, we report the observation of a dense triangular network of mid-gap 1D metallic modes in uniform and continuous ML $MoSe_2$ grown by molecular-beam epitaxy (MBE). Scanning tunneling microscopy and spectroscopy (STM/STS), high-resolution transmission electron microscopy (HRTEM) reveal these 1D modes are associated with inversion domain boundary (IDB) defects. Moreover, we find intensity undulations of these 1D modes in STM/STS micrographs and attribute to the superlattice potential of the moiré pattern and quantum confinement effect. The experiments of MBE growth, STM/STS, HRTEM, and density functional theory (DFT) calculations are presented in **Supplementary Materials S1-3** [33].

Figs. 1(a) and 1(b) show two STM images of the same surface under different bias conditions. The $MoSe_2$ film was grown on highly oriented pyrolytic graphite (HOPG) with the nominal thickness of 1.4 MLs. The terrace-and-step morphology and streaky reflection high-energy electron diffraction (RHEED) pattern (Fig. 1(a) and inset) suggest the layer-by-layer growth mode of MoSe$_2$. The spacing between the diffraction streaks matches with the lattice constant of a strain-free MoSe$_2$ crystal, in accordance with the van der Waals (vdW) epitaxy process [44]. From Fig. 1(a), one observes that the most part of the sample is of ML MoSe$_2$, but there are also holes of exposed substrate and the 2$^{nd}$ and 3$^{rd}$ layer MoSe$_2$ islands due to the > 1 ML nominal coverage and the kinetics of the MBE process. In Fig. 1(b) and the close-up images of Figs. 1(c) and 1(d), one also notes strikingly the wagon-wheel like patterns covering the entire surface of the sample. The bright "spokes" or rims are seen to compose of twin lines and remarkably these twin lines often exhibit intensity undulations (*cf*. Fig. 1(d)).

Set aside the intensity undulation, similar wagon-wheel like patterns were previously reported in multi-layer thick MoSe$_2$ films grown on MoS$_2$ and SnS$_2$ and were interpreted as the moiré interference effect [45,46]. Our analyses of the patterns in Fig. 1 do not conform to the moiré effect (see **Supplementary Materials S4**), however. As a matter of fact, we also observed the moiré patterns (e.g., the circled region in Fig. 1(e)) that are distinct from the wagon-wheels. One notes further that the moiré aligns with the rim, *i.e.*, along the zigzag or $\langle 11\bar{2}0 \rangle$ direction of MoSe$_2$ lattice.

By comparing Figs. 1(a) and 1(b), we suggest the bright contrasts of the wagon-wheel rims in the STM images are not topographic in origin but represents additional DOS confined in 1D, as their contrasts show STM-bias dependence and can sometimes disappear under certain conditions. To elucidate this further, we performed STS measurements and Fig. 2(a) shows the STS spectra taken far from (i) or in the vicinity of (ii) a rim. Comparing the two, one clearly observes extra peaks in (ii). Fig. 2(b) shows a STS map (more are found in **Supplementary Materials S5**), revealing the electronic modes confined in 1D and interconnected into the triangular network. Intensity undulations are also apparent. As will be discussed later, these metallic 1D modes are similar to the edge states of $MoS_2$ nanoclusters [20-25] and so they can be utilized for HDS and HER similar to nanoclusters.

In passing, we remark that spectra (i) shows semiconductor characteristics of ML $MoSe_2$ with the energy bandgap of ~ 2.1 eV. The optically measured exciton energy in ML $MoSe_2$ was ~ 1.6 eV [11,47,48], so our measurement would imply an exciton binding energy of ~ 0.5 eV, which is consistent with the first principles calculations [49-51].

To uncover the structural identity of the wagon-wheel pattern, we performed HRTEM experiments and Fig. 3(a) shows a Fourier-filtered HRTEM image (*cf.* **Supplementary Materials S2(b)**) of a sample grown under the same condition as Fig. 1 but at a smaller coverage of ~ 1 ML. The wagon-wheel structure is again discerned in HRTEM micrograph. The bright rims in STM images now manifest as the darker boundaries separating domains of triangular lattices of the bright spots. In Fig. 3(a), one observes rows of the bright spots that are misaligned or spatially shifted upon crossing the boundaries, which suggest "faults" in crystal. The fault and unfaulted domains (for ML $MoSe_2$, it is arbitrary to assign which domain is faulted) are delineated by domain boundaries. In Figs. 3(b) and 3(c), we show two close-up images revealing two typical structures of the domain boundaries. They separate domains of mirror symmetry and thus are likely inversion domain boundaries. These IDBs are essentially intersections of two $MoSe_2$ grains with a relative in-plane rotation of $180^o$, which effectively swaps positions of Mo and Se lattice sites. In ML

MoSe$_2$, inversion domains may be created by stacking faults of either Mo or Se [34,52]. A number of possible atomic configurations exist for such a defect, some of which are shown in **Supplementary Materials S2(b)**. The model of Fig. 3(d), in which the surface Se atoms are faulted while the middle layer Mo atoms maintain their closed packed stacking, most likely reflects the defects in the MBE-grown MoSe$_2$ according to our first principle calculations and combinational STM and TEM measurements (*cf.* **Supplementary Materials S2 and S6**). TEM simulations based on the model of Fig. 3(d) reproduce the experimental results well (see the overlaid images in Figs. 3(b) and 3(c) and the superimposed ball models) if one or two rows of Se atoms at the IDB (*i.e.*, those pointed by the short blue arrows in Fig. 3(d)) are vacant. The latter is a reasonable assumption due to high-energy (200 keV) electron irradiation of the sample during TEM experiments [38]. In fact, in the image of Fig. 3(a), one sees not only the two types of the boundary structures but also a mixture of the two for one and the same boundary (*e.g.*, the one encircled by box '3' in Fig. 3(a)), implying incomplete desorption of Se atoms at the defects during TEM experiment.

While domain boundary appears common in monolayer TMDCs, they have been mostly seen in isolation or irregularly distributed in exfoliated flakes or CVD-grown samples [31,35,36,38]. It is striking to see them interconnected into a regular network in the MBE-grown film. A network of the 1D metallic modes expectedly bring about important consequences to the properties of the films, particularly the catalytic and transport behaviors, which would require future experimental and theoretical attention.

First principle calculations of the model of Fig. 3(d) result in the DOS and band dispersion as shown in Figs. 2(c) and 2(d), respectively. Qualitatively, the main features are in agreement with the experiment (Fig. 2(a)) but there are also discrepancies. There might be lattice stretching/compression in epitaxial MoSe$_2$ in the vicinity of defects as well as some kind of reconstruction of the edge atoms, which are not explicitly considered in our calculations. The offset in energy of the defect states between theory and experiment is caused by Fermi energy difference. In calculation, a ribbon-model was used and the Fermi energy was in the middle of the energy bandgap, whereas for the MBE sample, the Fermi level was close to the conduction band edge due to unintentional doping by defects. The twin lines of the defect in STM and STS images reflect the spatial distribution of the DOS, which has been reproduced in the calculated DOS maps as exemplified in Figs. 2(e) and 2(f). In the atomically resolved STS and DOS maps (*cf*. Fig. 2(f)), the bright contrasts are seen not to coincide with Se or Mo atomic positions but in the interstitial region of Mo-Se honeycombs. This is found similar to the edge states of MoS$_2$ clusters [19]. The mid-gap states induced by the IDBs in ML MoSe$_2$ are also similar to the edge states of ML nanoclusters, so they may find similar applications such as for HDS and HER [20-25].

As noted earlier, another and perhaps more interesting feature seen in Figs. 1(c), 2(b) and 4(a,b) is the intensity undulations along the bright rims. These undulations are not inherent to the IDB defects themselves, as our calculated DOS maps for a free-standing MoSe$_2$ with the IDB never show such undulations. We suggest they are caused by the coupling between MoSe$_2$ and HOPG substrate. In order to obtain more quantitative information about the undulations, we extract the intensity data and carry out least-square fittings using a squared sinusoidal function. Examples are shown in the inset of Fig. 4(c) and the derived wavevectors are summarized in the main panel of Fig. 4(c) for a total of seven IDB defects with different lengths. As is seen, there are mainly three energy ranges in which the intensity undulations are prominent. In all three, the most apparent undulations have the same wavelength of $\lambda$ ~ 1 nm (corresponds to a wavevector of ~ 0.32 Å$^{-1}$), coinciding with the moiré periodicity in the system (examples are shown in the top panel of **Supplementary Material S5**). Our STM examinations suggest crystallographic corrugations cannot be the cause for such intensity undulations. The interaction between MoSe$_2$ overlayer and HOPG substrate depends on atomic registry within the moiré pattern, which can have two effects. First, the STM/STS maps can in principle directly reflect such position dependent interaction between the epilayer and the substrate. Second, the moiré pattern realizes an effective superlattice potential that modifies the behavior of conducting electrons as recently demonstrated in graphene [53-55]. In our experiment, the strong undulations at the moiré periodicity are seen to occur prominently over three energy intervals as marked by the dotted circles in Fig. 4(c) rather than continuously over the wide energy range in the bulk bandgap. Therefore, we believe the second effect is more relevant here, namely the moiré pattern has induced a superlattice potential along the line of the defect and consequently standing waves emerged due to Bragg reflection of the moiré superlattice potential. In other words, the moiré superlattice potential may have led to band folding of the electronic states in a reduced Brillouin zone (BZ) as is exemplified in Fig. 4(d). At the reduced BZ boundary, the dispersion relations are modified (*e.g.*, the gap opening) such that states at the reduced BZ boundary become "Bragg-reflected". The three energies at which intensity undulations are observed at the same wavelength of 1 nm may then correspond to Bragg reflected waves of the different branches of the electronic bands as marked by the blue arrows on the right of Fig. 4(d), for example.

Besides undulations at the same wavelength of the moiré period, we also observe another set of undulations over the energy range from -0.9 to -0.5 eV (the bottom panel of **Supplementary Materials S5**), where the undulation wavelength *depends* on energy (*cf.* the middle encircled set of data in Fig. 4(c)). This corresponds to a dispersive dependence of the wavevector on energy and we attribute it to standing waves of the quantum well states (QWS) due to finite lengths of the defects that are pierced by the intersections with other defects in the network. The intersection junctions serve as the boundaries or scattering centers of electron waves. The QWS is evidenced by the finding that for a given IDB, the neighboring undulation wavelengths satisfy approximately the relation $\lambda/\lambda' = n/(n+1)$, where *n* is an integer. From the energy range such dispersive undulations are observed and their off-core spatial distribution, we assign them to correspond to states of the highlighted branch of the dispersion in Fig. 4(d). We remark however that due to unattainable atomic resolution STS images in this energy range, a comparison with calculated DOS maps remains ambiguous, so the above assignment is preliminary. It shows the points of argument, however. An interesting point about these undulations is that the derived wavevectors, as fitted by the single sinusoidal function $I \sim \sin^2(kx + \varphi)$, are confined to values of $\leqslant 0.32$ Å$^{-1}$, *i.e.*, within the first reduced BZ of the moiré superlattice potential.

Periodic (moiré) potentials confined in 1D in a finite length represents an example of the classic "electrons-in-finite-length-superlattice" problems, where standing waves exhibit multiple components with wavevectors $\pm k + n\mathrm{G}$, where $\mathrm{G}$ is the reciprocal lattice vector of the superlattice, and the amplitudes of the components drop with *n* [56-58]. The noise of the data does not allow us to perform the Fourier transform to reliably get all these amplitudes. Nevertheless, since the standing waves in such systems have the dominating component of $n = 0$ [58], fitting by a single sinusoidal function is justifiable. And the fact that the extracted wavevectors are all less than 0.32 Å$^{-1}$ is consistent with the band-folding picture by the moiré superlattice potential as postulated above.

To summarize, we have grown ML MoSe$_2$ on HOPG by MBE and found a dense network of mid-gap 1D electronic modes. The high density and high DOS of these 1D modes will suggest great potentials of such continuous TMD films for catalysis applications. In addition, spatially undulated intensities along the bright rims in STM/STS images are recorded, which are attributed to moiré superlattice potential and the quantum confinement effect. The interconnection of the 1D conducting channels into triangular network expectedly causes important electronic, optical and catalytic responses that may impact on future applications of the films.

**Acknowledgements**

We are benefited from discussions with X.D. Cui, C.H. Liu, B. Li, J.L. Chen and K.Q. Hong. Y.F. Chan helped in some early TEM examinations of the sample. MHX, NW and WY acknowledge the support of the CRF grant (No. HKU9/CRF/13G) from the Research Grant Council of Hong Kong Special Administrative Region, China. MHX and JFJ also acknowledge the support from the MoE/RGC joint research scheme (No. M-HKU709-12). CLG and JFJ acknowledge supports from National Basic Research Program of China (Nos. 2012CB927401, 2011CB921902, 2011CB922200, and 2013CB921902), and the NSFC Grants (Nos. 11374206, 91021002, 11274228, 10904090, 11174199, and 11134008). WY is also supported by Research Grant Council Grant (No. HKU705513P).

**Figures and Captions:**

**FIG. 1 STM micrographs of MoSe$_2$-on-HOPG**. (**a**) Topographic image (size: 80×80 nm$^2$, sample bias $V_{sample}$ = 2.0 V) of a 1.4 ML MoSe$_2$ film. The white arrows point to the exposed substrate ('1') and the 2$^{nd}$ ('2') and 3$^{rd}$ ('3') layer MoSe$_2$ islands, respectively. The inset is the RHEED pattern taken along [11$\bar{2}$0]. (b) Topographic image of the same area as (a) but at $V_{sample}$ = -1.0 V. (**c, d**) Close-up STM images (size: 35×35 nm$^2$, $V_{sample}$ = -1.0 V for (c); 13×13 nm$^2$ and $V_{sample}$ = -1.46 V for (d)), revealing the wagon-wheel patterns and the intensity undulations. (**e**) STM image (size: 9.2×9.2 nm$^2$, $V_{sample}$ = -1.78 V) showing the moiré pattern, e.g., in the upper-left corner of the micrograph. The inset is the intensity profile along the white line, showing the period of the moiré pattern to be of ~ 1 nm.

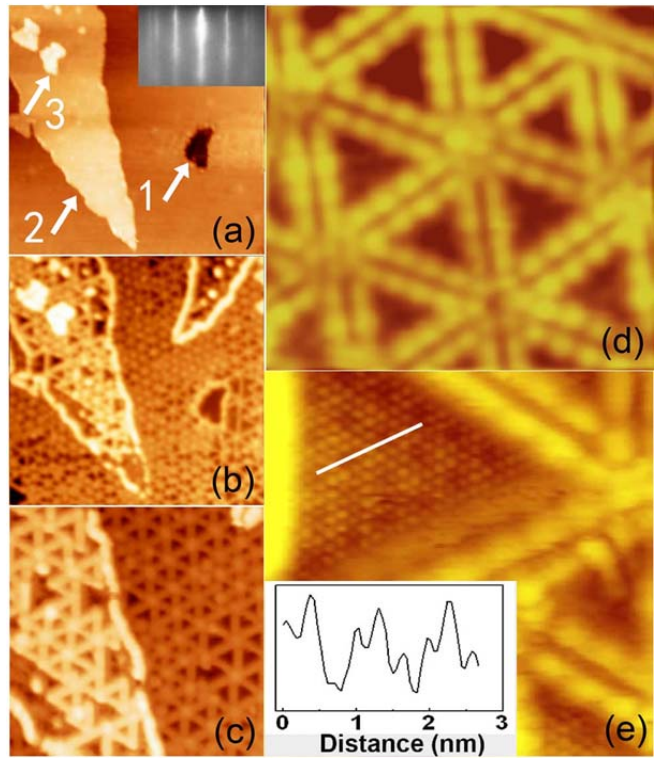

**FIG. 2 STS spectra and calculated DOS of the IDB defect**. (**a**) STS spectra of ML MoSe$_2$ at the points as marked by the crosses in (b), i.e., away from (i) or in the vicinity of (ii) a defect. (**b**) A STS map (size: 8.6×8.6 nm$^2$) obtained at $V_{sample}$ = -0.42 eV, showing the 1D defect modes and their intensity undulations along the lengths of the defects. (**c, d**) Calculated DOS and the band structure. The red and green dots in (d) represent states from Mo and Se atoms at the core of the defect, respectively, and the blue dotted states are from Mo next to the core. In (c), the red-shaded regions are the valence and conduction bands of bulk ML MoSe$_2$ and the black line is the DOS from the defect. (**e, f**) STS maps obtained at $V_{sample}$ = -1.85 eV and 0.72 eV, respectively, superimposed with the calculated DOS maps and the ball model of the structure (refer to Fig. 3(d) below) at the corresponding energies. The calculated maps represent the central region of a larger ribbon model (*cf.* **Supplementary Material S3**), showing the DOS of the defect.

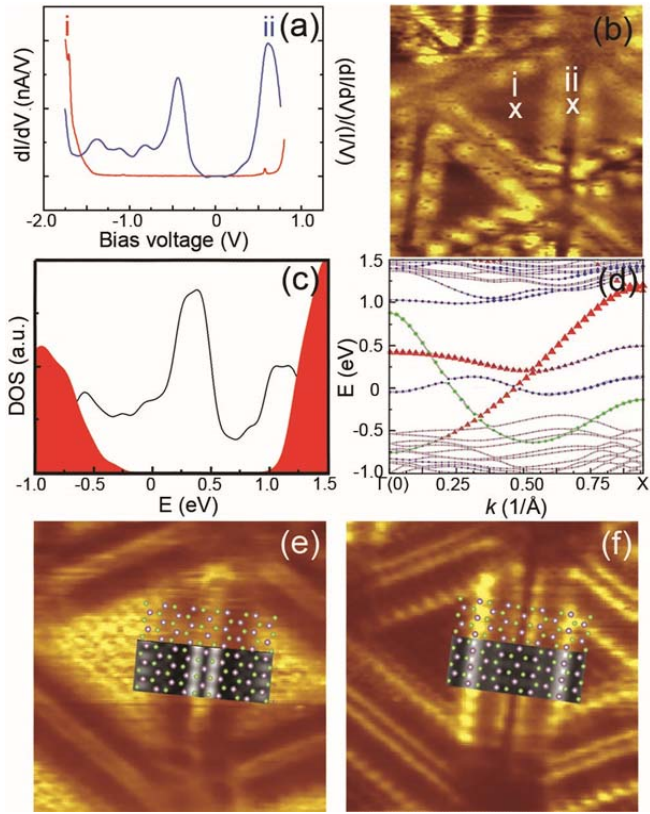

**FIG. 3 HRTEM micrographs of the defect** (**a**) HRTEM image of ML MoSe$_2$ on HOPG(graphene). The red/yellow lines mark rows of the bright spots in two neighboring domains. The numbered boxes highlight the boundaries corresponding to one ('1') or two ('2') Se rows vacant, or a mixture (box '3') of the two structures. (**b, c**) Simulated HRTEM images (the highlighted regions) based on the model of (d) but with one or two rows of Se atoms removed at the defect as shown by the overlaid models (the purple and green balls represent Mo and Se atoms, respectively). (**d**) Stick-and-ball model of the IDB defect in both plan-view (top) and cross-sectional view (bottom).

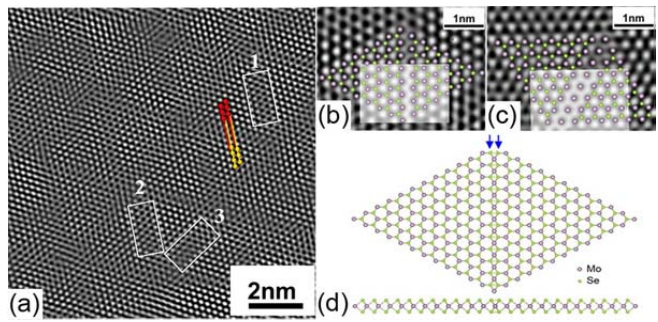

**FIG. 4 STS maps and extracted intensity undulations of the 1D modes.** (**a, b**) STS maps (size: 4.5×4.5 nm$^2$) at energies as indicated. The intensities $I$ along the lengths of the defects (marked by the white lines) are extracted and fitted by equation $I = A\sin^2(kx + \varphi) + D$, where $A$, $k$, $\varphi$ and $D$ are fitting parameters, as exemplified in the **inset** of (c). (**c**) Wavevectors derived by fittings of seven defects of different lengths (labeled by different colors) and at different energies. The dotted and solid circles highlight the undulation wavevectors that correspond to the moiré period and the QWS, respectively. (**d**) Energy bands of Fig. 2(d) but folded into the reduced BZ of the moiré supperaltice potential.

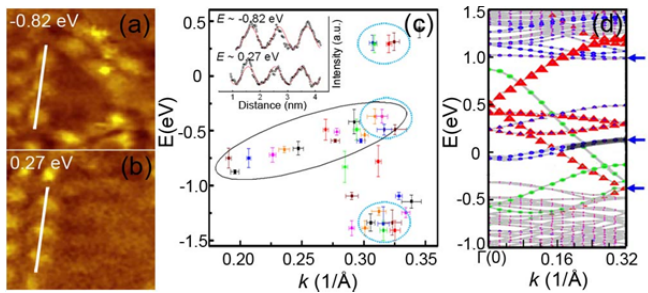